\documentclass[12pt]{iopart}
\usepackage{amsmath}
\usepackage{graphicx}
\usepackage{siunitx}
\usepackage{color}
\usepackage[normalem]{ulem}
\usepackage[backend=biber,url=false,doi=false,isbn=false]{biblatex}
\usepackage[acronym, shortcuts]{glossaries}
\newacronym{SERF}{SERF}{spin-exchange relaxation free}
\newacronym{MOT}{MOT}{magneto-optical trap}
\newacronym{LAE}{LAE}{laser-assisted etching}
\newacronym{ZFR}{ZFR}{zero-field resonance}
\newacronym{KOH}{KOH}{potassium hydroxide}
\newacronym{DRIE}{DRIE}{deep reactive ion etching}

\begin{document}

\title{Microfabricated multi-axis cell for integrated atomic devices}

\author{L. Péroux$^1$, A. Dewilde$^1$, A. Mursa$^2$, A. Mazzamurro$^1$, J. Bonhomme$^1$, Q. Tanguy$^2$, E. Klinger$^2$,  L. Gauthier-Manuel$^2$, O. Gaiffe$^2$, A. Talbi$^1$, R. Boudot$^2$, P. Pernod$^1$, J.-F. Clément$^1$, V. Maurice$^1$ and N. Passilly$^2$}
\address{$^1$ Univ. Lille, CNRS, Centrale Lille, Univ. Polytechnique Hauts-de-France, UMR 8520 - IEMN - Institut d’Electronique de Microélectronique et de Nanotechnologie, Lille, France}
\address{$^2$ Université Marie et Louis Pasteur, CNRS, SUPMICROTECH, institut FEMTO-ST, Besançon, France}
\ead{linda.peroux@centralelille.fr, nicolas.passilly@femto-st.fr}

\vspace{2pc}

\begin{abstract}
Microfabricated alkali vapor cells enable the miniaturization of atomic sensors, but require collective wafer-level integration of complex features. In many applications—including magnetometers, gyroscopes, magneto-optical traps, and fluorescence imaging—multiple optical accesses are needed to enhance performance. Yet, achieving this without compromising manufacturability remains challenging. In this work, we present a wafer-level fabrication approach that enables three orthogonal optical pathways in microfabricated alkali vapor cells, using fully scalable and collective processes. Our method relies on the thermal reflow of glass preforms, shaped by laser-assisted etching (LAE) and bonded between silicon frames. The relatively low surface roughness produced by LAE allows for effective reflow, which further smooths the surfaces without significantly compromising the optical planarity of the windows. This process results in multi-axis vapor cells featuring embedded, optics-grade lateral windows. We evaluate the device performance through both single-beam and dual-beam atomic magnetometry measurements. Magnetic sensitivities better than \SI{200}{\femto\tesla\per\sqrt\hertz} are demonstrated along each of the three orthogonal axes, confirming the potential of the approach for tri-axis magnetic field sensing at sub-picotesla resolution. This fabrication strategy opens new perspectives for versatile, high-performance atomic sensors, fully compatible with wafer-level integration and mass production.
\end{abstract}

%
%
%
%
%

\section{Introduction}

Atomic devices enable highly precise measurements of time, magnetic field or angular velocity. However, their widespread use has been hindered by the cost, size, and scalability of their components. In particular, these devices rely on alkali vapor cells, which were traditionally produced through glassblowing, a process dependent on craftsmanship, making it unsuitable for miniaturization and mass production.
In the early 2000s, researchers at NIST pioneered microfabrication techniques for alkali vapor cells, addressing these limitations~\cite{liewMicrofabricatedAlkaliAtom2004}.
Originally developed for the semiconductor industry, microfabrication enables precise structuring of materials at the microscale, allowing the production of compact and low-power atomic systems~\cite{kitchingChipscaleAtomicDevices2018}. 
The first microfabricated alkali cells quickly led to advancements in chip-scale atomic clocks~\cite{knappeMicrofabricatedAtomicClock2004} and magnetometers~\cite{schwindtChipscaleAtomicMagnetometer2004a}, demonstrating their potential for miniaturized quantum technologies.
Beyond miniaturization, microfabrication offers additional advantages, in particular the ability to co-integrate microelectrical features (such as heaters and thermistors) and micro-optics, while sustaining greater manufacturability~\cite{raghavanFunctionalizedMillimeterscaleVapor2024a}.
However, most microfabricated vapor cells are designed with a single optical access, which restricts the use of multi-beam configurations that could enhance their performance or enable additional functionalities.
For instance, centimeter-scale glass-blown cells with two optical accesses have demonstrated magnetometry sensitivities as low as \SI{0.6}{\femto\tesla\per\sqrt{\hertz}}~\cite{kominisSubfemtoteslaMultichannelAtomic2003}. 
Concerning microfabricated cells, sub-picotesla sensitivity was demonstrated in \ac{SERF} regime but the performance was limited by amplitude noise which could have been suppressed using a second probe beam~\cite{shahSubpicoteslaAtomicMagnetometry2007}. One solution involves tilting the cell to allow orthogonal pump and probe beams, leading to \SI{5}{\femto\tesla\per\sqrt\hertz} sensitivity at \SI{200}{\celsius}~\cite{griffithFemtoteslaAtomicMagnetometry2010}. An alternative solution uses superimposed beams, which has demonstrated sub-picotesla scalar magnetometry~\cite{zhangSubpicoteslaScalarAtomic2019,gerginovScalarMagnetometry1002020}.
Furthermore, multiple beam configurations offer additional advantages, for instance by enabling \ac{MOT}, which has been demonstrated on microfabricated cells with a 6-beams configuration enabled by the use of a large window~\cite{boudotEnhancedObservationTime2020}.
To further enhance compactness microfabricated grating chips have been combined with microfabricated cells to realise \ac{MOT} with a single beam but larger volumes or additional transverse optical access may be needed~\cite{mcgilliganLaserCoolingChipscale2020}.
Similarly, terahertz imaging systems based on atomic vapor require two separate optical paths: one for delivering the pump lasers that excite the atoms, and another for collecting the visible fluorescence used to form the image~\cite{downesFullFieldTerahertzImaging2020}.
Alternative designs have been explored to overcome these limitations. Spherical glass-blown cells on wafers, developed for gyroscopes~\cite{eklundGlassblownSphericalMicrocells2008}, allow multiple optical accesses. However, the non-uniformity of the wall thickness results in distortion of the transmitted optical beams through the glass-blown structure, restricting its usability in certain applications. Other approaches exploit integrated reflective surfaces to enable compact optical routing in microfabricated cells~\cite{chutaniLaserLightRouting2015a,perezRubidiumVaporCell2009a}.
Femtosecond laser machining enabled the fabrication of cells with customizable 3D shapes, which can improve optical access and has already been demonstrated in a single-beam magnetometry measurement~\cite{luciveroLaserwrittenVaporCells2022,zanoniPicoteslaOpticallyPumped2024}.
More recently, microfabricated atomic vapor cells with multiple optical channels have been demonstrated using an inner-sidewall molding process~\cite{yuMicrofabricatedAtomicVapor2023}. While this approach improves optical access, its complexity introduces significant fabrication challenges. Here, we propose a microfabrication process to realise cells with three perpendicular high quality optical accesses at the wafer-level. 

\section{Microfabricated multi-axis cell}

\subsection{Concept}

\begin{figure}[ht!]
\centering
\includegraphics[width=15cm]{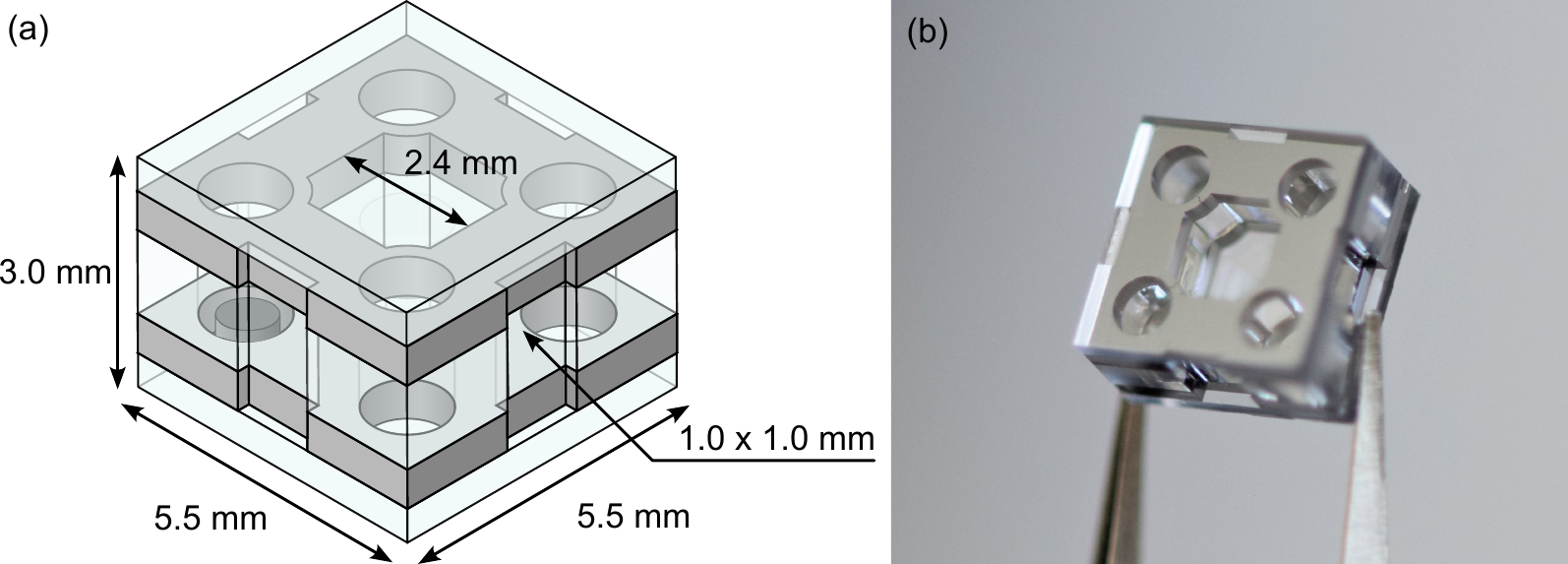}
\caption{(a) Cell schematic, (b) cell photography. }
\label{fig:1}
\end{figure}

The multi-axis cell is fabricated from a stack of five wafers: three glass wafers and two silicon wafers (Fig.~\ref{fig:1}). This configuration is designed to incorporate a central glass wafer that can be microstructured to provide lateral optical windows. The central glass is bonded between two external glass wafers using intermediate silicon wafers, enabling robust and hermetic anodic bonding—a technique well-established for alkali vapor cell fabrication \cite{kitchingChipscaleAtomicDevices2018}.
This five-wafer structure is employed to achieve higher depths than what can be obtained with deep reactive ion etching, as in Ref. ~\cite{petremandMicrofabricatedRubidiumVapour2012a}.
In our case, the two intermediate silicon wafers also serve as a mechanical frame during the glass reflow process, which is carried out at the wafer-level on the Si–glass–Si sandwich alone. This rigid frame helps maintain the geometry of the etched glass cavities, preventing deformation or collapse during reflow and thereby preserving the flatness of the optical windows. The central glass wafer is structured using laser-assisted wet etching, a technique that offers several advantages. Compared to direct laser ablation, it produces smoother surfaces and vertical (90°) sidewalls, requiring only minor surface correction during reflow to achieve optical quality windows. Additional cavities are etched along the dicing lines to form recessed walls, which undergo the same glass-reflow processing as the main cavities. These recessed features help preserve the optical quality of the outer surfaces of the windows during wafer dicing, thereby eliminating the need for post-polishing. This approach also allows the use of a coarser dicing blade, significantly improving the efficiency of cell release.

\subsection{Fabrication}

\begin{figure}[ht!]
\centering
\includegraphics[width=15cm]{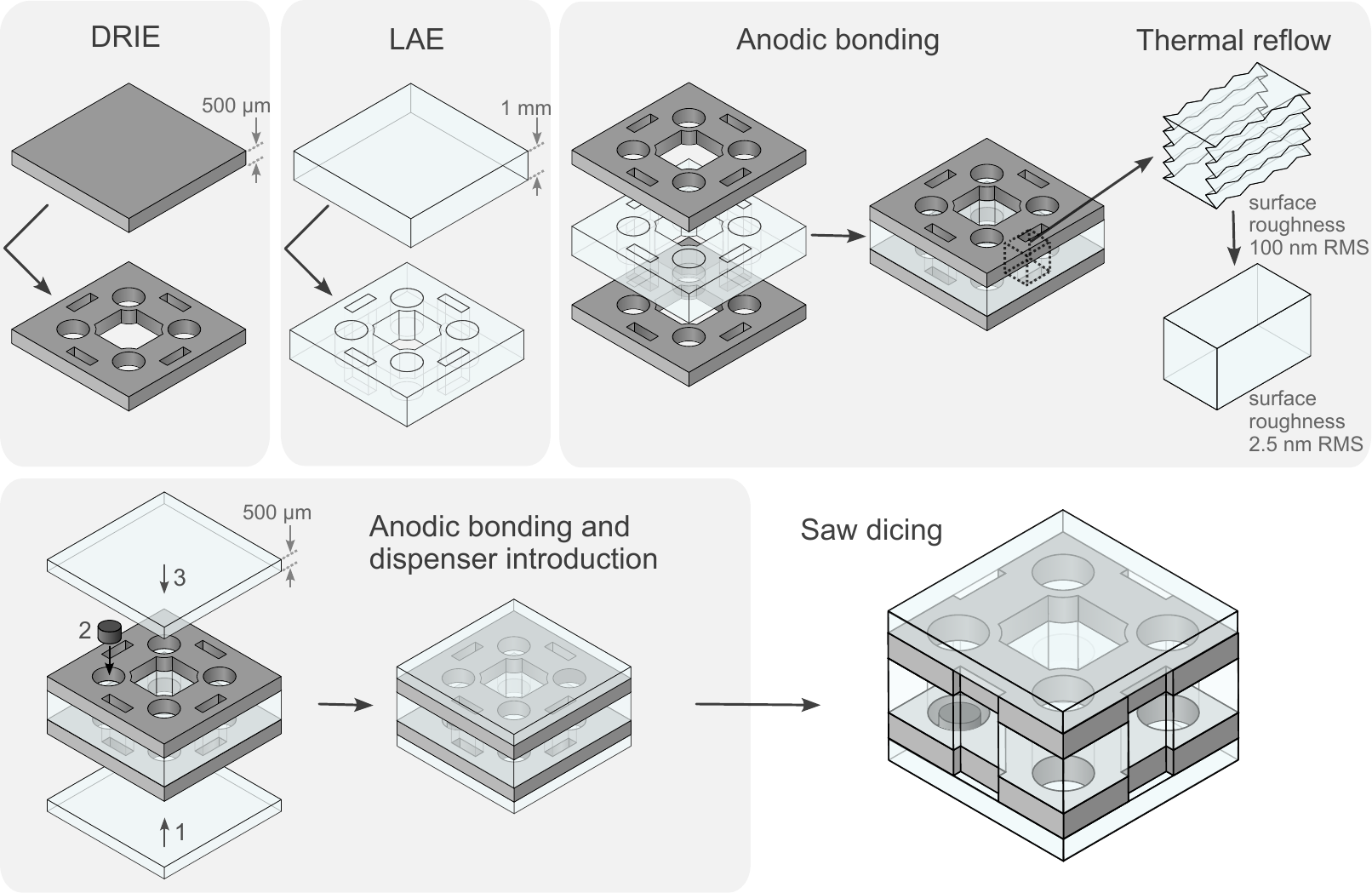}
\caption{Microfabrication flowchart: The central glass substrate, structured by \ac{LAE}, is sandwiched between two silicon wafers, used first to create a rigid frame during thermal reflow, and secondly, which allow relying on anodic bonding for the cell sealing. External cavities prevent contact with the saw blade during cell release.}
\label{fig:flowchart}
\end{figure}

The fabrication process, shown in Fig.~\ref{fig:flowchart}, begins with the structuring of both the silicon and the central glass wafers. The silicon wafers, \SI{500}{\micro\meter} thick, are processed by optical lithography and \ac{DRIE} to form cavities that are slightly smaller than those in the glass wafer. This deliberate offset ensures proper edge anchoring of the glass cavities during the subsequent thermal reflow step. These silicon wafers also include narrow channels that connect the dispenser cavity to the science chamber.
The \SI{1}{\milli\meter}-thick central borosilicate glass wafer (BOROFLOAT33, SCHOTT, Germany) is structured using \ac{LAE}. Unlike laser ablation, this method involves direct laser writing only to define the cavity contours, which are then developed by wet etching. By doing so, the
sacrificial material does not require to be entirely laser-irradiated but is eventually detached during
wet etching in a \ac{KOH} aqueous solution. Although \ac{KOH} is primarily known for anisotropic etching of silicon, its etching rate is significantly enhanced in laser-exposed regions of glass, enabling 3D structuring. Originally proposed in the early 2000s, femtosecond laser pre-irradiation was first used to enhance etching in hydrofluoric acid solutions~\cite{marcinkeviciusApplicationBesselBeams2001}, but higher selectivity was later demonstrated with \ac{KOH}~\cite{matsuoLaserInternalModification2009}.
The laser writing step is carried out using a f100 aHead Enhanced system (FEMTOprint, Switzerland), where the glass wafer is mounted horizontally on an XY translation stage. This stage is positioned above a vertically movable 50× microscope objective, which focuses a \SI{1030}{\nano\meter} femtosecond Yb:YAG laser beam into the glass bulk. The \ac{LAE} process relies on localized structural modifications induced by the intense laser irradiation at the focal point.
Etching follows in a 10~mol/L \ac{KOH} solution, maintained at 85$^{\circ}$C under intermittent ultrasonic agitation. The laser-modified regions are etched hundreds of times faster than unexposed glass~\cite{rossOptimisationUltrafastLaser2018}, allowing selective removal with minimal impact on surrounding areas. Unlike laser ablation, the resulting cavity edges remain sharp and clean, enabling bonding without requiring additional surface treatments such as polishing.
After etching, the lateral walls of the cavities typically exhibit a surface roughness around \SI{100}{\nano\meter} RMS—significantly smoother than what is usually obtained by laser ablation~\cite{yuMicrofabricatedAtomicVapor2023}, yet still insufficient for optical-quality surfaces. To improve surface quality, a thermal reflow is performed. The reflow takes place in vacuum (typically a few times $10^{-6}$~\SI{}{\milli\bar}), where the temperature is ramped at 6$^{\circ}$C/min up to 850$^{\circ}$C (slightly above the glass softening point). The surface is smoothened for 20 minutes before a controlled cooldown. A 30-minute annealing step at 560$^{\circ}$C is added to relieve internal stresses.
This reflow process is conducted on a Si-glass-Si stack, pre-assembled by anodic bonding performed at 350$^{\circ}$C and 900~V. The silicon wafers act as a mechanical frame, ensuring structural integrity during reflow and preventing the glass corners from collapsing. The silicon cavities are designed to be nearly \SI{50}{\micro\meter} narrower than their glass counterparts, to reinforce edge anchoring and accommodate potential bonding misalignments. 
Final sealing is achieved by anodically bonding two \SI{500}{\micro\meter} thick borosilicate glass wafers onto both sides of the reflowed structure: Once the first external wafer is bonded, the alkali dispensers are loaded into their designated cavity. The second glass wafer is then bonded in a controlled atmosphere (neon or nitrogen) to complete the cell sealing. Alkali dispensers are then laser-activated at the wafer level. After confirming the presence of saturated alkali vapor through optical absorption measurements, the cells are released from the wafer by means of saw-dicing.

\subsection{Windows characterisation}

Unlike DRIE, where sidewall roughness of silicon cavities is typically characterized by scalloping due to alternating etch/passivation cycles, \ac{LAE} reveals a different form of periodic surface modulation. In our case, the vertical sidewalls exhibit a fine, regular texture which corresponds to the successive translations of the laser focal point, layered sequentially in depth during the irradiation process. The pattern then arises from the layered voxel-based exposure strategy and is later revealed by differential etching rates between laser-modified and unmodified glass regions.
As a result, the surface quality is orientation-dependent, with smoother walls typically observed along directions aligned with the laser scanning path. In the vertical direction, the roughness is generally about 50$\%$ higher than in the horizontal direction. 

In the following, surface topography and roughness have been measured using an optical profilometer based on white-light interferometry (MSA 500 from Polytec, Waldbronn, Germany), over an  $\SI{886}~\times~\SI{662}{\micro\meter}^2$ area corresponding to the field of view of a $\SI{10}~\times$ Mirau objective. As mentioned earlier, it is noteworthy that the sidewall roughness resulting from wet etching is relatively low compared to surfaces produced by direct laser ablation ~\cite{yuMicrofabricatedAtomicVapor2023}. Additionally, within \ac{LAE}-processed structures, roughness of the sidewalls is slightly lower than that observed at the cavity bottom as reported in Ref.~\cite{skoraHighfidelityGlassMicroaxicons2022}, despite the much larger spacing between parallel laser tracks (from \SI{1}{\micro\meter} overlap to \SI{10}{\micro\meter} separation). This difference stems directly from the ovoid shape of the voxel. In details, Ref.~\cite{skoraHighfidelityGlassMicroaxicons2022} reported roughness values of $R_q~=~$\SI{45}{\nano\meter} RMS along the writing lines and $R_q~=~$\SI{90}{\nano\meter} RMS perpendicular to them, whereas here, we observe a roughness of $R_q~=~$\SI{47}{\nano\meter} RMS along the writing lines and approximately $R_q~=~$\SI{64}{\nano\meter} RMS in the vertical direction.

Nevertheless, in their current state, these surfaces are not smooth enough to transmit laser beams without introducing distortions, making thermal reflow necessary. Surface annealing is therefore carried out in a tubular furnace (HST1200 by Carbolite) under vacuum conditions. To calibrate the reflow temperature, a Si-glass-Si structured stack was first fabricated, and diced into chips, which were then annealed at various setpoint temperatures. For each chip, the setpoint temperature was maintained for 20 minutes, followed by controlled cooling at 5$^{\circ}$C/min down to room temperature.

\begin{figure}
\centering
\includegraphics[width=15cm]{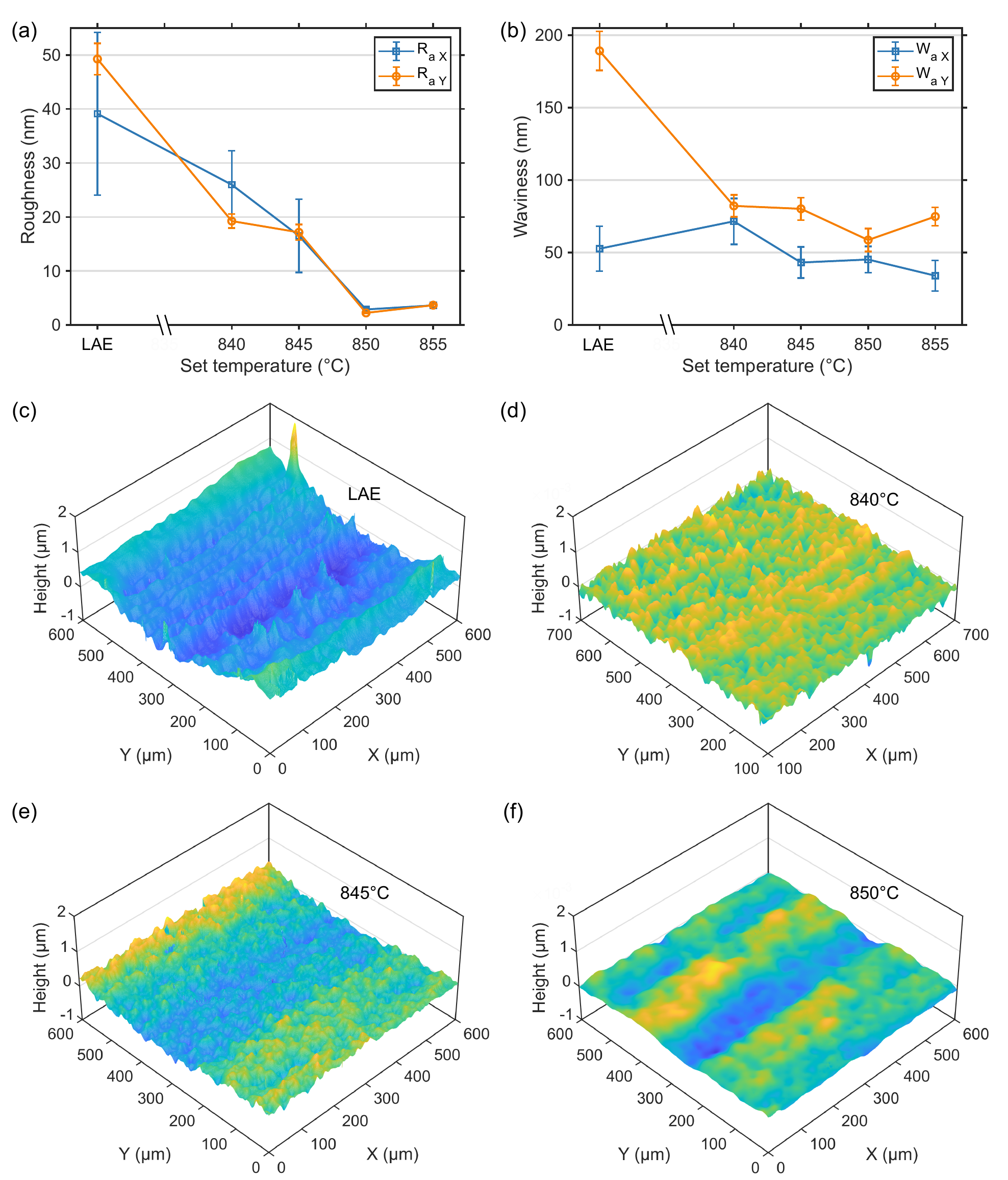}
\caption{Improvement of cell window surface with thermal annealing: (a) Surface roughness and (b) surface waviness as a function of the maximal reflow temperature, surface topography of samples (c) before and (d-f) after thermal reflow.}
\label{fig:Roughness} 
\end{figure}

The evolution of surface quality for different chips as a function of the reflow temperature is shown in Fig.~\ref{fig:Roughness}. Surface roughness decreases significantly as the annealing temperature approaches 850$^{\circ}$C, until nearly optical quality is reached. Anisotropy in surface roughness persists after reflow, but becomes much less pronounced at higher annealing temperatures. Around 850$^{\circ}$C, this anisotropy almost vanishes, and the surface roughness $R_a$ falls below \SI{3}{\nano\meter} in all directions.

It is worth noting that the surface parameters of the samples are strongly modified by reflow, although the temperature increments are small. Combined with slight variations in chip dimensions (all cut from the same wafer), this can explain the variability observed in the curves. Nevertheless, the overall trend is clear: the initially high vertical waviness decreases rapidly, while horizontal waviness is more affected by the surface reorganisation at first (840$^{\circ}$C). Then, as the reflow temperature increases, roughness along both directions is found to be smaller and more uniform. At higher temperatures (e.g. above 850$^{\circ}$C), however, surface bowing increases, indicating that a tradeoff must be found between minimizing roughness and preserving surface flatness.

In the following, the inner stack of the fabricated cells was reflowed at the wafer-level, with the annealing temperature adjusted accordingly.

\subsection{Cells functionalized for magnetometry}

\begin{figure}
\centering
\includegraphics[width=15cm]{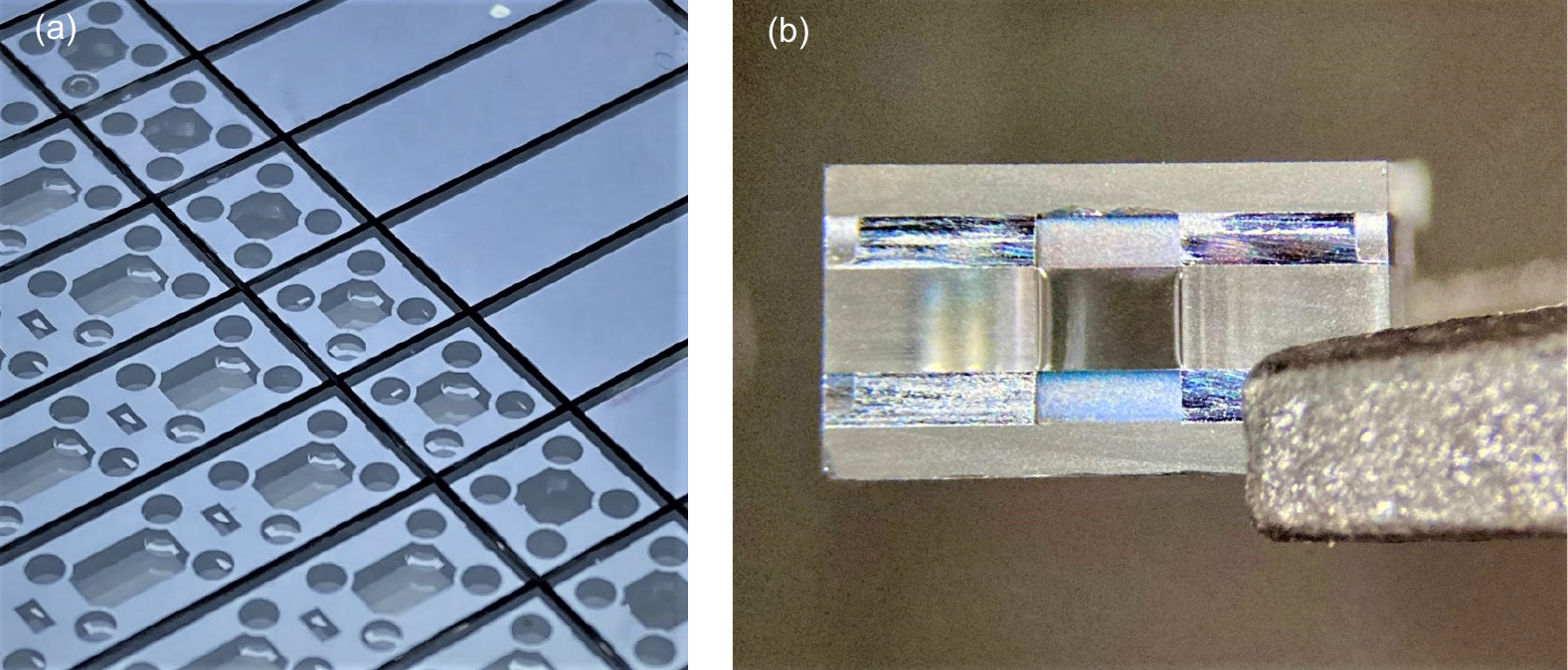}
\caption{(a) Wafer after saw dicing. (b) Cell lateral side and window.}
\label{fig:FabCells} 
\end{figure}

Once the Si-glass-Si stack was thermally reflowed, it was assembled with the bottom glass lid and loaded with Cs dispensers. The top glass lid was then anodically bonded to seal the cell under an atmosphere of 760~Torr of nitrogen at \SI{350}{\celsius}. As expected, the subsequent laser activation of the dispenser consumes a portion of the nitrogen~\cite{dyerNitrogenBufferGas2023a}. Indeed, whereas a pressure of nearly 380~Torr could be expected without dispenser, the nitrogen pressure, determined by measuring the induced frequency shift, is found to be 330~Torr at \SI{60}{\celsius}. The wafer right after saw dicing is shown on the Fig.~\ref{fig:FabCells}(a). This step produces rough external sidewalls, except at the recessed windows, as illustrated in Fig.~\ref{fig:FabCells}(b).

\section{Magnetometry measurements}


To evidence the cell potential, a \ac{ZFR} magnetometer is implemented using one of the fabricated multi-axes cell filled with nitrogen. In this setup, two distributed Bragg reflector (DBR) lasers are used to generate the pump and probe beams along the z and x-axis respectively. The pump beam is tuned to the $^{133}\mathrm{Cs}$ $D_1$ line (895 nm) and the probe beam is sligthly detuned from the $^{133}\mathrm{Cs}$ $D_2$ line (852 nm), both lasers are free-running.
The cell is placed in a three-layer magnetic shield, and the magnetic field is established using a set of three coils. Measurements are carried out by measuring the change in absorption of the transmitted pump light or the polarization state rotation of the transmitted probe light. The pump beam is collimated and passes through a half-wave plate followed by a quarter-wave plate to produce circular polarization. The probe beam is linearly polarized and its amplitude is adjusted using a half-wave plate. After passing through the cell, the probe beam is split by a polarizing beam splitter (PBS) into two orthogonally polarized components, which are directed onto a differential photodiode. The intensity difference measured between these two beams provides a signal proportional to the polarization rotation. 

The light intensities are estimated by measuring the optical power transmitted through the cold cell and considering the window size. The cell temperature and laser intensities are tuned manually in order to optimize sensitivity. To evaluate the magnetometer sensitivity, the noise spectrum is calibrated by generating oscillating magnetic fields of known amplitude at several frequencies between 10 and \SI{100}{\hertz}. The measurement is taken from either the lock-in amplifier output (for the one-beam configuration) or the differential photodiode output (for the two-beam configuration).

\begin{figure}
\includegraphics[width=15cm]{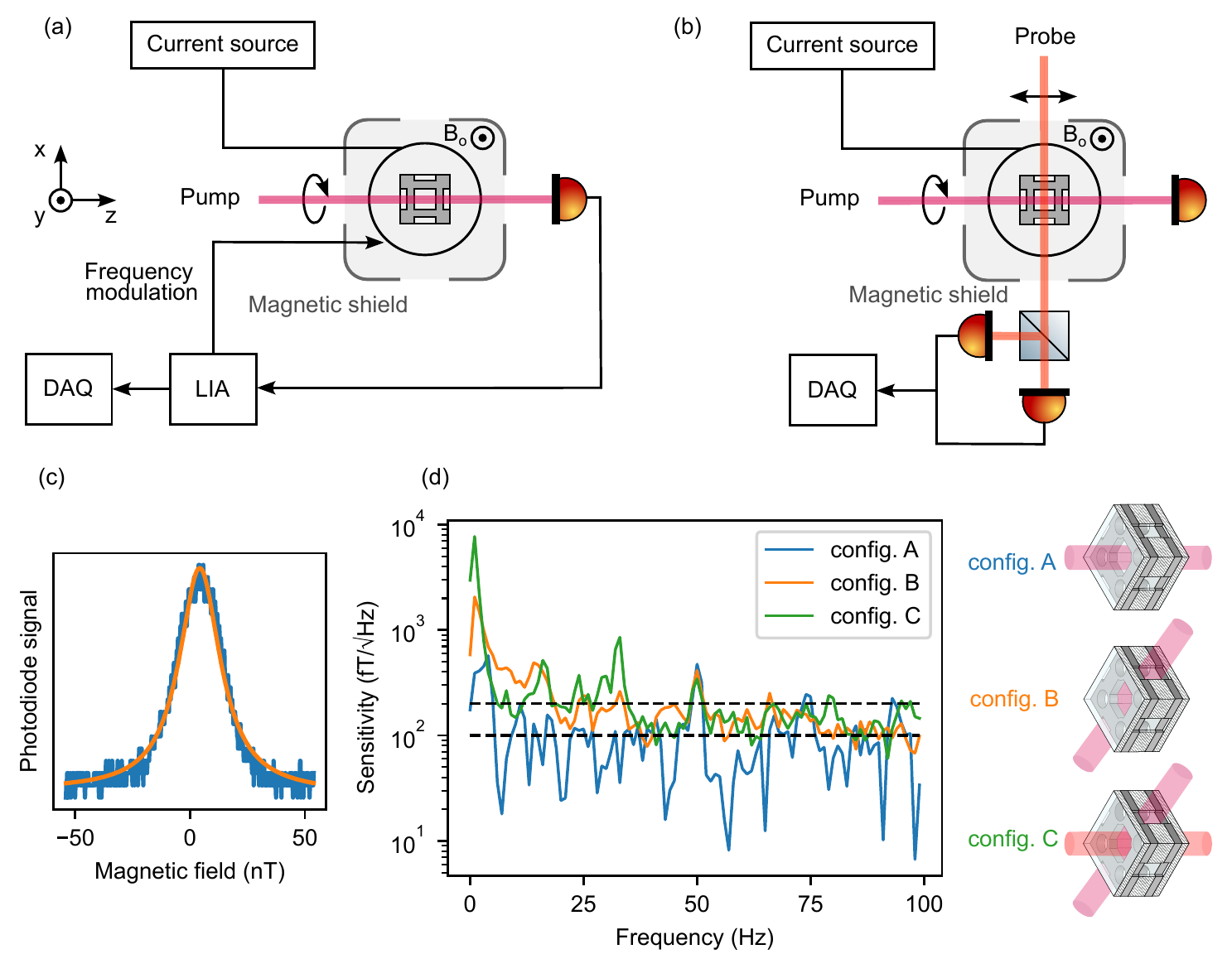}
\caption{\label{fig:SERF_AA} Experimental setup of the \ac{ZFR} magnetometer for measurements: (a) in absorption and (b) in polarization rotation. (c) Zero-field resonance signal measured in transmission by scanning the magnetic field $B_0$ (config. B). The orange line indicates a Lorentzian fit with a full-width at half-maximum of \SI{22}{\nano\tesla} at \SI{104}{\celsius} with a \SI{12}{\milli\watt\per\centi\meter\squared} laser intensity. (d) PSD (Power Spectral Density) of the magnetometer signal for the different configurations. The dotted lines indicate \SI{100}{\femto\tesla\per\sqrt{\hertz}} and \SI{200}{\femto\tesla\per\sqrt{\hertz}}.}
\end{figure}

Magnetometry measurements are carried out using two different setups, shown in Fig.~\ref{fig:SERF_AA}(a) and (b). The first setup, based on a single-beam geometry, relies on absorption detection whereas the two-beam geometry detects the polarization rotation of the probe beam. By scanning the magnetic field around zero along the y-axis, a zero-field resonance is observed (Fig.~\ref{fig:SERF_AA}(c)). For the measurements performed with a single beam, the magnetic field along the y-axis is modulated and the dispersive signal is obtained by demodulating the photodiode signal with a lock-in amplifier.
To compare the response along different cell axes, absorption measurements are performed in two configurations : either with the pump light passing through the glass lids (configuration A) or through the lateral windows (configuration B). In configuration A, the pump light intensity is \SI{1.6}{\milli\watt\per\centi\meter\squared} and the cell is operated at a temperature of \SI{76}{\celsius}. In configuration B, the pump light intensity is \SI{14}{\milli\watt\per\centi\meter\squared} and the cell is operated at \SI{90}{\celsius}.
In the two-beam configuration (C), the magnetometer signal is obtained from polarization rotation of the probe beam. The pump light illuminates the vapor through the lateral windows with an intensity of \SI{14}{\milli\watt\per\centi\meter\squared} while the probe light uses the main windows with an intensity of \SI{12.6}{\milli\watt\per\centi\meter\squared}. In this configuration, the cell is operated at a temperature of \SI{99}{\celsius}. The Power Spectral Density (PSD) of the magnetometer signal for each configuration is shown in Fig.~\ref{fig:SERF_AA}(d). In configuration A, the sensitivity noise floor reaches \SI{100}{\femto\tesla\per\sqrt\hertz} while in configurations B and C it reaches \SI{200}{\femto\tesla\per\sqrt\hertz}.

\section{Discussion and conclusion}

A part of the sensitivity difference between the configurations can be attributed to the smaller illuminated atomic volume, resulting from the reduced dimensions of the lateral windows, as well as their slightly lower optical quality. In configuration A, the probed volume is $\SI{2.4}{\milli\meter} \times \SI{2.4}{\milli\meter} \times \SI{2.0}{\milli\meter} = \SI{11.5}{\milli\meter\cubed}$, while in configurations B and C, it is $ \SI{1.4}{\milli\meter} \times \SI{1.0}{\milli\meter} \times \SI{2.4}{\milli\meter} = \SI{3.4}{\milli\meter\cubed}.$ Since sensitivity scales inversely with the square root of the probed volume, this volume ratio of 3.4 corresponds to a sensitivity ratio of approximately 1.8 under comparable experimental conditions. In addition, diffraction at the edges of the lateral window, combined with its lower flatness, may alter the polarization state. Increasing the dimensions of the lateral window and/or shaping the probe beam could therefore enable sensitivities in configurations B and C comparable to those achieved in configuration A. Furthermore, the buffer gas pressure is currently limited by the maximum pressure the bonding machine can withstand, but increasing this pressure could likely improve sensitivity. An alternative approach could be to fill the cell externally~\cite{perouxLocallySealedMicrofabricated2025}, eliminating the need for internal dispensers while allowing higher buffer gas pressures.

Compared to prior approaches~\cite{yuMicrofabricatedAtomicVapor2023}, our method offers significant advantages. By employing \ac{LAE} rather than direct laser ablation, we achieve an intrinsically lower level of roughness on the lateral glass windows. This eliminates the need for a mold to emboss the glass surface since a simple thermal reflow improves the surface roughness to $R_a~<~$\SI{3}{\nano\meter}. Consequently, issues related to alignment, mold wear, and dimensional accuracy are avoided, enabling simpler and more reliable mass production while improving consistency across large-scale manufacturing.
Moreover, the introduction of auxiliary cavities ensures that the saw blade never comes in contact with the windows, thereby eliminating the need for an additional polishing step.
Despite the limited buffer gas pressure of 330~Torr, the fabricated cells achieved sensitivities of \SI{100}{\femto\tesla\per\sqrt\hertz} with optical access through the main window and \SI{200}{\femto\tesla\per\sqrt\hertz} through the lateral windows.

\section*{Data availability}
The data that support the findings of this study are available from the corresponding author upon reasonable request.

\section*{Acknowledgments}
This work was supported by the Direction Générale de l'Armement (DGA) and by the Agence Nationale de la Recherche (ANR) in the frame of the ASTRID project named NOMAD (Grant ANR-23-ASTR-0020) and EIPHI Graduate school (Grant ANR-17-EURE-0002). The PhD thesis of L. Péroux is co-funded by Agence Innovation Defense (AID) and Région Hauts-de-France. The PhD thesis of A. Dewilde is funded by the ANR (Grant ANR-23-CE42-0012).
This research work has been partially undertaken with the support of IEMN
and FEMTO-ST microfabrication facilities (CMNF and MIMENTO),
both part of the French Renatech network.

\printbibliography

\end{document}